\pdfoutput=1
\documentclass[conference,letterpaper]{IEEEtran}
\IEEEoverridecommandlockouts
\newif\ifanonymous
\anonymousfalse 

\usepackage{cite}
\usepackage{amsmath}
\usepackage{graphicx}
\usepackage{booktabs,tabularx,array,multirow}
\usepackage{url}

\usepackage{newtxtext,newtxmath}

\graphicspath{{figures/}}
\newcolumntype{Y}{>{\raggedright\arraybackslash}X}
\newcolumntype{P}[1]{>{\raggedright\arraybackslash}p{#1}}

\begin{document}
\bstctlcite{IEEEexample:BSTcontrol}

\title{Bilateral Protection for Private LoRA Inference with Intel TDX}

\ifanonymous
\author{\IEEEauthorblockN{Anonymous Authors}
\IEEEauthorblockA{Anonymous Affiliation}}
\else
\author{\IEEEauthorblockN{Xiaodong Li, Mingyuan Ma\textsuperscript{*}, and Dongbo Yan}
\IEEEauthorblockA{Beijing Electronic Science and Technology Institute, Beijing 100070, China\\
Email: lxd@besti.edu.cn, maming\_yuan@163.com, ydb362522@163.com}
\thanks{\textsuperscript{*}Corresponding author.}}
\fi

\maketitle
\thispagestyle{empty}

\begin{abstract}
Cross-organizational AI collaboration increasingly combines a public base language model with a task-specific low-rank adaptation (LoRA) module owned by an independent specialist to process a client organization's confidential inputs on shared cloud infrastructure. Because an untrusted cloud operator hosts execution and neither stakeholder wishes to expose its asset to the other, both the client and the adapter owner must verify that the exact same isolated service receives their secrets and executes the approved base model, tokenizer, and adapter. We design and implement a bilateral confidential inference service inside an Intel Trust Domain Extensions (TDX) trust domain (TD) in which each party independently verifies the hardware Quote, replayed boot measurements, and its Transport Layer Security (TLS) handshake public key before releasing the prompt or the weight-decryption key, while an in-memory loader passes verified model and decrypted adapter tensors directly to the inference engine without reopening filesystem paths. Across three model sizes (0.6B, 4B, and 8B parameters), the protected service reproduces same-TD native outputs token for token on 400 MultiWOZ dialogue states, 100 DialogSum summaries, and 32 MeetingBank agenda-item minutes, adding a geometric-mean remote-to-native request-time ratio of 1.0632 across 12 independent boots of the 0.6B service.
\end{abstract}


\section{Introduction}
Cross-organizational workflows often apply specialized domain models to confidential operational records without centralizing data ownership. Low-rank adaptation (LoRA) supports this division of labor by freezing pretrained base weights and encoding domain expertise in compact low-rank update matrices~\cite{lora}. An untrusted cloud operator can thus host a public base model while pairing a specialist's private adapter with a client's sensitive prompts, provided both stakeholders can verify that the same isolated service receives their secrets, applies the intended adapter to the agreed base model and tokenizer, and returns outputs solely to the client.

Hardware trusted execution environments (TEEs) isolate plaintext computation from untrusted hosts, while attested Transport Layer Security (TLS) binds encrypted channels to hardware evidence~\cite{ratls}. Using these primitives, SeSeMI mediates user and model policies through an attested key service for serverless enclaves~\cite{sesemi}, LoRATEE keeps private adapters inside a TEE while offloading base linear layers to an untrusted GPU~\cite{loratee}, and OpenPCC binds client session keys to attested language-model nodes~\cite{openpcc}.

\begin{figure*}[t]
\centering
\includegraphics[width=396pt]{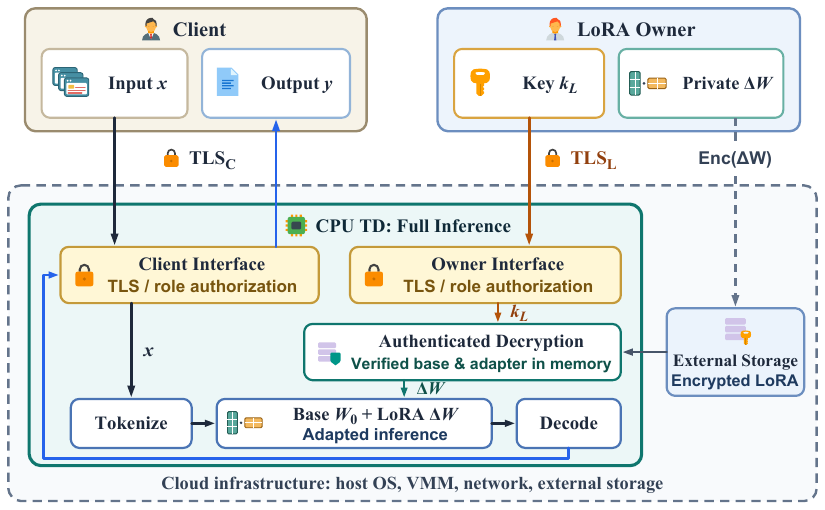}
\caption{Three-party cooperative inference architecture. The client and LoRA owner independently verify the CPU TD service and its TLS public-key binding over $\mathrm{TLS}_C$ and $\mathrm{TLS}_L$ before releasing $x$ and $k_L$. TLS termination, authenticated decryption, and full inference execute inside the TD.}\label{fig:arch}
\end{figure*}

Cooperating across an independent client and LoRA owner requires extending trust from TLS termination to the runtime objects actually executed. First, unilateral attestation or a valid Quote alone does not prove that both parties approved the same measured application on the live TLS endpoints they reach. Second, checking file digests at a filesystem path fails if a mutable path changes before the SDK reopens it, while public base weights and tokenizers jointly govern prompt interpretation. Third, loading structurally valid LoRA tensors has no effect if attachment or runtime scaling leaves their contribution at zero.

We build a bilateral single-adapter LoRA service inside an Intel Trust Domain Extensions (TDX) trust domain (TD). Over separate TLS~1.3 sessions, the client and LoRA owner independently verify fresh Quotes, UEFI Secure Boot and boot-event logs, and their TLS handshake public keys before releasing the decryption key or prompts; tokenization, authenticated decryption, base-plus-LoRA execution, and decoding run entirely inside the CPU TD. Our contributions are:
\begin{enumerate}
\item \emph{Independent bilateral service admission:} Separate client and owner verifiers enforce UEFI Secure Boot and strict \texttt{UpToDate} TDX and Quoting Enclave (QE) policies on Google Cloud TDX VMs, rejecting an unapproved boot-menu update before key release.
\item \emph{Direct verified-byte loading and activation checks:} Our loader constructs OpenVINO model, tokenizer, and AES-256-GCM decrypted LoRA objects in memory without reopening paths, validated via package-tampering tests and 4B/8B activation ablations.
\item \emph{Decoupled fidelity, wrapper, and cross-device evaluation:} Across Qwen3-0.6B, Qwen3-4B, and Meta-Llama-3.1-8B-Instruct, we test paired token preservation on three benchmarks, isolate steady-state wrapper and Secure Boot startup costs, and benchmark matched \texttt{llama.cpp} generation across TD CPU, TITAN RTX, and an attested CC-On NVIDIA H20 GPU.
\end{enumerate}

\section{Related Work}
\textbf{Attested Execution and Confidential Inference.}
Attested TLS binds channel keys to TEE evidence~\cite{ratls}, while Parma, Confidential Containers, and Gramine tie attested workloads to launch policies or manifest-checked files~\cite{parma,coco,gramine}. Prior confidential inference systems vary in isolation scope: MLCapsule encloses local enclave execution~\cite{mlcapsule}, Slalom offloads linear layers via blinding and verification~\cite{slalom}, SLIM obfuscates split LLM execution across TEEs and untrusted GPUs~\cite{slim}, and Chrapek et al.\ quantify full-model CPU vs.\ GPU TEE costs~\cite{confllm}, motivating in-memory consumption checks and runtime-matched baselines.

\textbf{Multi-Adapter Serving and Bilateral Governance.}
S-LoRA, Punica, and dLoRA batch many adapters on trusted GPU stacks~\cite{slora,punica,dlora}, whereas cross-organizational private adapters must bind two stakeholders' secrets to a jointly approved base and adapter. Unlike third-party key brokers or split CPU--GPU pipelines (Table~\ref{tab:related})~\cite{sesemi,loratee,openpcc}, our single-TD CPU design lets both stakeholders directly attest the same endpoint and binds their approvals to verified in-memory runtime objects.

\begin{table}[t]
\caption{Comparison of Confidential Inference and Private-Adapter Systems}\label{tab:related}
\scriptsize
\renewcommand{\arraystretch}{1.05}
\begin{tabularx}{\columnwidth}{P{1.48cm}YYY}
\toprule
\textbf{System} & \textbf{Authorization} & \textbf{Execution Scope} & \textbf{Evaluation Focus} \\
\midrule
SeSeMI~\cite{sesemi} & KeyService mediates user/model policy & Serverless enclave runtime & Serverless inference overhead \\
LoRATEE~\cite{loratee} & Multi-tenant adapter service & Adapter in TEE; base linear on GPU & Heterogeneous multi-LoRA speed \\
OpenPCC~\cite{openpcc} & Client verifies node \& session keys & Confidential LLM serving node & Prompt privacy \& serving cost \\
\textbf{This work} & Bilateral client \& owner endpoint verification & Verified in-memory base, tokenizer, \& LoRA in one CPU TD & Strict hardware admission, paired fidelity, wrapper vs.\ GPU-TEE speed \\
\bottomrule
\end{tabularx}
\end{table}

\section{Bilateral System Design}
\subsection{System, Threat, and Hardware Admission Model}
The system coordinates a client holding confidential prompt $x$ (receiving $y$), a LoRA owner holding encrypted adapter $\Delta W$ and key $k_L$, and an untrusted cloud operator supplying the TDX VM, network, and external storage with $\Delta W$, public base weights $W_0$, and tokenizer files (Fig.~\ref{fig:arch}). Both parties pre-approve $W_0$, its tokenizer, and configuration $\theta$ governing
\begin{equation}
y=\operatorname{Generate}(W_0,\Delta W,x;\theta),\qquad
\Delta W=(\alpha/r)BA,
\label{eq:lora}
\end{equation}
with rank $r$ and scale $\alpha$. Role separation blocks clients from $\Delta W$ or $k_L$ and isolates $x$, $y$, and client TLS keys from the owner.

\begin{figure*}[t]
\centering
\includegraphics[width=396pt]{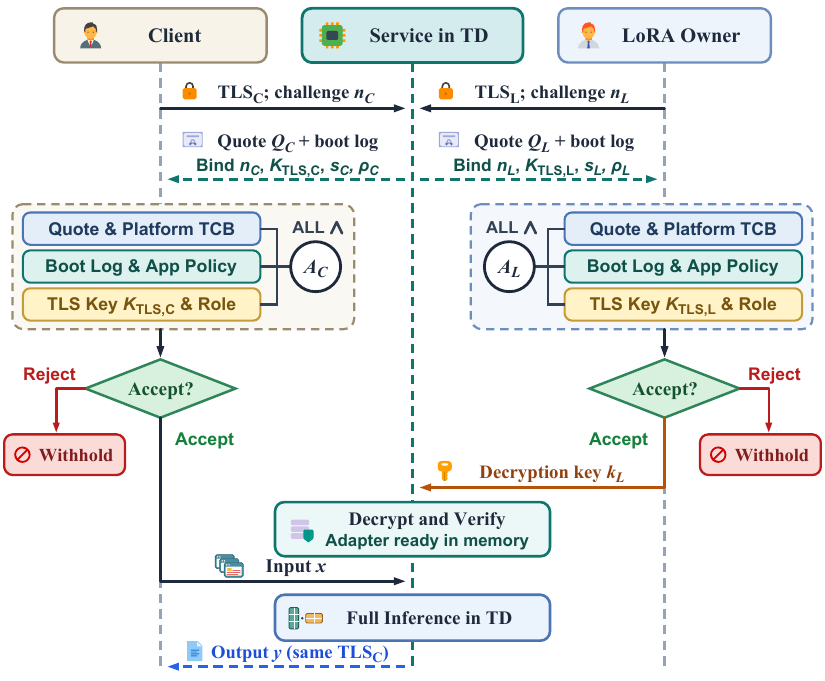}
\caption{Bilateral remote attestation and secret release. Subscripts $C$ and $L$ denote client and LoRA owner. Each party independently evaluates platform TCB state, boot/application policy, and handshake-key binding via conjunctive appraisal ($\mathcal{A}_C$, $\mathcal{A}_L$ in~\eqref{eq:accept}) on its own TLS session before releasing $k_L$ or $x$.}\label{fig:protocol}
\end{figure*}

An adversary controls the host OS, VMM, storage, and network and may alter stored files, replay Quotes, or redirect traffic. Both parties trust Intel TDX hardware and root certificates and approve the measured guest TCB (firmware, GRUB, kernel, service binary, and OpenVINO SDK) to enforce: (i)~\emph{secret confidentiality} in TD memory and role channels; (ii)~\emph{execution-object integrity} over $\Delta W$, $W_0$, tokenizers, and scaling flags; and (iii)~\emph{bilateral release control}. Both verifiers require UEFI Secure Boot~\cite{gcpshieldboot} and \texttt{UpToDate} TDX platform and QE status against current Intel recovery collateral~\cite{inteltcbrecovery}, withholding secrets on unpatched or outdated hosts until a compliant instance re-attests (including after Google Cloud C3 maintenance restarts~\cite{gcpmaintenance}).

\subsection{Bilateral Attestation and Role-Bound Secret Release}
Acting as independent relying parties~\cite{rats}, the client and owner open separate TLS~1.3 sessions~\cite{tls}, send fresh 32-byte challenges $n$, and extract the server SubjectPublicKeyInfo (SPKI) key $K_{\mathrm{TLS}}$~\cite{ratls} (Fig.~\ref{fig:protocol}). Keeping its TLS private key in TD memory, the service writes $d=\operatorname{SHA512}(c)$ into the TDX Quote's 64-byte \texttt{REPORTDATA}~\cite{tdx}:
\begin{equation}
c=\operatorname{Canon}(p,n,K_{\mathrm{TLS}},s,\rho,w,H(L)),\quad d=\operatorname{SHA512}(c),
\label{eq:binding}
\end{equation}
binding protocol version $p$, challenge $n$, key $K_{\mathrm{TLS}}$, 32-byte session ID $s$, role $\rho\in\{\mathrm{C},\mathrm{L}\}$, workload ID $w$, and SHA-256 digest $H(L)$ of the Confidential Computing Event Log (CCEL).

For role $\rho$ with policy $\pi_\rho$, secret release requires
\begin{equation}
\begin{aligned}
\mathcal{A}_\rho={}&\operatorname{CryptoValid}(Q)\land\operatorname{PlatformAllowed}(Q,\pi_\rho)\\
&\land\operatorname{ApplicationAllowed}(Q,L,w,\pi_\rho)\\
&\land\operatorname{BindingMatches}(Q,c).
\end{aligned}\label{eq:accept}
\end{equation}
Upon accepting $\mathcal{A}_\rho$, each party signs $(d,\text{purpose})$ with its role credential; the service verifies this signature before enabling key delivery on $\mathrm{TLS}_L$ or inference on $\mathrm{TLS}_C$ under bounded session lifetimes and monotonic request counters. Over $\mathrm{TLS}_L$, the owner sends 256-bit key $k_L$; inside TD memory, the service decrypts $\Delta W$ via AES-256-GCM~\cite{gcm} (binding manifest version, adapter ID, and plaintext SHA-256 as additional authenticated data), checks tensor structure, loads the pipeline, marks readiness, and locks $k_L$. Clients then send $x$ and receive $y$ solely over $\mathrm{TLS}_C$.

\subsection{Immutable Guest, Verified-Byte Loading, and Activation}
The service binary, OpenVINO runtime, and manifest reside in a read-only SquashFS root inside a measured \texttt{initramfs}; verifiers check \texttt{MRTD}, replay CCEL events against \texttt{RTMR}s, and match reference digests (guest GRUB inspection confirmed Linux boots the exact measured \texttt{initramfs} buffer). The worker runs unprivileged with dropped capabilities and \texttt{no\_new\_privs}, with swap, core dumps, consoles, guest agents, module loading, \texttt{kexec}, and \texttt{ptrace} disabled; diagnostic flags exist only in separately measured test images.

To eliminate time-of-check-to-time-of-use (TOCTOU) path swaps, our loader enforces $H(b)=h_b$ and $b_{\mathrm{SDK}}=b_{\mathrm{verified}}$ in memory without reopening paths across: (i)~base XML and INT4 weights (re-hashing NumPy/OpenVINO \texttt{Tensor} byte views); (ii)~tokenizer/detokenizer buffers; (iii)~JSON configs; and (iv)~AES-256-GCM decrypted LoRA tensors (never written to disk). Before OpenVINO GenAI attachment~\cite{openvino}, a contract checks projection names, shapes $A_j\in\mathbb{R}^{r\times d_{\mathrm{in}}}$, $B_j\in\mathbb{R}^{d_{\mathrm{out}}\times r}$, types, rank $r$, and scale $\lambda=\alpha/r$ in $f_j(h)=W_{0,j}h+\lambda B_j(A_j h)$ ($r=16,\alpha=32\Rightarrow\lambda=2$; 392 tensors across 28 layers in Qwen3-0.6B), complemented by \emph{enabled} ($\lambda=2$), \emph{disabled}, and \emph{zero-scale} ($\lambda=0$) CPU ablations.

\section{Experimental Methodology}
\subsection{Cloud TDX Deployments, Models, and Paired Tasks}
We deployed on two Singapore Google Cloud C3 TDX VMs under Secure Boot and \texttt{UpToDate} TDX/QE policies: \texttt{4-vCPU, 16~GB} (0.6B replication and 12-boot cost study) and \texttt{8-vCPU, 32~GB} (three-model scale and summarization studies). Protected and same-TD native workers use OpenVINO INT4 bases, four threads, one stream, batch size one, and greedy decoding with Qwen thinking disabled~\cite{qwen3} (verified-byte loading on all 32~GB runs). An off-cloud controller runs both verifiers and times requests; a TITAN RTX workstation trained adapters ($r=16,\alpha=32$, dropout $0.05$, lr $2\times 10^{-4}$, batch 8) and ran FP16 PyTorch references. Across Qwen3-0.6B, Qwen3-4B, and Meta-Llama-3.1-8B-Instruct~\cite{llama3}, paired workers process identical chat-templated prompts (4096-char, 1024-input-token cap) on:
\begin{itemize}
\item \textbf{MultiWOZ~2.2~\cite{multiwoz}:} Trained on 3000 dialogues, selected on 240 dev turns by exact match (EM) then slot micro-F1 (epochs 7, 4, 9 on 32~GB; BF16 epoch-11 0.6B on 16~GB), and tested on 400 turns (200 hotel, 200 restaurant; 64-token cap, strict JSON parsing).
\item \textbf{DialogSum~\cite{dialogsum}:} Trained on 1200 dialogues, selected over two epochs on 80 dev dialogues by ROUGE-L F1~\cite{linrouge}, and tested on 100 dialogues (three references; 64-token cap).
\item \textbf{MeetingBank~\cite{meetingbank}:} Trained on 234 pairs, selected over three epochs on 12 dev meetings, and tested on 32 agenda items (512-new-token ceiling, natural EOS stopping).
\end{itemize}

\subsection{Matched Cross-Device Generation and Overhead Timers}
To avoid confounding hardware speed with runtime or quantization choices, a single \texttt{llama.cpp} binary ran identical \texttt{Q4\_K\_M} GGUF bases, unmerged FP32 LoRA exports, and 400 pre-tokenized MultiWOZ prompts (6 warm-ups, fresh KV cache, greedy, 64-token cap) on the 32~GB TD CPU, TITAN RTX (4B, 8B), and a Beijing NVIDIA H20 GPU (\texttt{ecs.gn8v-tee.4xlarge}) in CC-On mode with debugging disabled (8B, three runs). Before H20 runs, our verifier checked a fresh 32-byte challenge against the signed GPU report, certificate chain, and ES384 NVIDIA Remote Attestation Service (NRAS) verdict. Timers span prefill through the final token (excluding loading, tokenization, and network), using H20 as a device-attested generation baseline alongside Singapore CPU-TD bilateral release.

For OpenVINO steady-state overhead, we measure controller HTTPS \emph{remote}, worker \emph{internal}, and \emph{same-TD native} times over 6 warm-ups and 24 timed prompts ($\{64,256,768\}$ input $\times$ $\{16,64\}$ output tokens): 12 boots on 16~GB summarized by geometric mean $R=\exp(\frac{1}{12}\sum_{j=1}^{12}\log r_j)$ (5000-boot bootstrap 95\% CIs), and 3 boots per model on 32~GB (alternating arm order) logging \emph{key-receipt-to-readiness}, \emph{reboot-to-readiness}, peak RSS, and 250~ms guest memory (\texttt{MemTotal}$-$\texttt{MemAvailable}). Six alternating 32~GB boot pairs (\texttt{7.0.0-1017-gcp}, rank-16 0.6B) isolated Secure Boot startup cost.

\begin{table}[t]
\caption{Admission and Component Verification Evidence}\label{tab:security}
\scriptsize
\renewcommand{\arraystretch}{1.05}
\begin{tabularx}{\columnwidth}{P{1.82cm}P{2.02cm}Y}
\toprule
\textbf{Check} & \textbf{Environment} & \textbf{Observed Outcome} \\
\midrule
Bilateral admission & Singapore 16~GB TD (MultiWOZ) & Both roles verified fresh Quotes \& TLS keys across 144 sessions (12 boots) \\
Boot-policy refusal & Singapore 32~GB TD & Approved task images admitted; 3 changed GRUB events rejected pre-$k_L$ \\
Package integrity & Local controller receiver & 4/4 mismatches (ID, digest, key, size) rejected per model before readiness \\
Adapter activation & Local CPU (4B \& 8B, 2 dev prompts) & Enabled ($\lambda=2$) altered outputs; disabled \& zero-scale ($\lambda=0$) matched \\
\bottomrule
\end{tabularx}
\end{table}

\section{Experimental Evaluation}
\subsection{Bilateral Admission, Artifact, and Activation Evidence}
All Singapore task runs passed bilateral admission under Secure Boot and \texttt{UpToDate} TDX/QE policies, including 144 verified sessions across 12 boots on the 16~GB TD (Table~\ref{tab:security}). Prior to the 8B matched-device run on 32~GB, an automatic guest update altered three GRUB entries; despite \texttt{UpToDate} TDX/QE status, the verifier rejected the boot log before $k_L$ release (confirmed against a verified maintenance boot). Local receiver tests rejected all 4/4 injected package mismatches (ID, digest, key, size) per model, while 4B/8B CPU ablations confirmed $\lambda=2$ altered outputs and $\lambda=0$ matched disabled runs token for token.

\subsection{Output Preservation, Accuracy, and Cross-Device Speed}
Protected and same-TD native workers matched token for token on 400/400 MultiWOZ turns on 16~GB (60.00\% EM, 91.51\% micro-F1) and on all 400 MultiWOZ turns plus 72 cost-grid pairs per model on 32~GB, reaching 60.25\%, 64.25\%, and 66.50\% EM (Table~\ref{tab:scale}; TITAN RTX FP16 references: +41.75~pp to 64.75\% at 4B, +48.75~pp to 65.75\% at 8B).

\begin{figure}[t]
\centering
\includegraphics[width=\columnwidth]{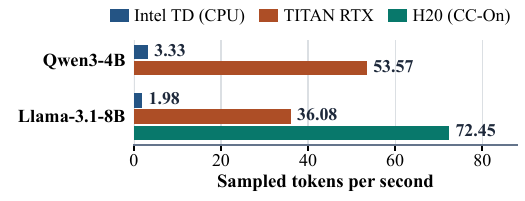}
\caption{Matched batch-one generation throughput across 400 MultiWOZ prompts using a shared \texttt{llama.cpp} worker, \texttt{Q4\_K\_M} bases, and FP32 LoRA. The 8B H20 result pools three fresh worker runs in CC-On mode.}\label{fig:throughput}
\end{figure}

In matched \texttt{llama.cpp} (\texttt{Q4\_K\_M}) generation across 400 MultiWOZ prompts (Fig.~\ref{fig:throughput}), the CC-On H20 reached 72.45 tokens/s (0.414~s mean generation, 0.130~s TTFT) at 8B vs.\ 36.08 tokens/s (0.829~s) on TITAN RTX and 1.98 tokens/s (15.115~s) on TD CPU---$36.60\times$ TD CPU and $2.01\times$ TITAN RTX speedups (4B: 53.57 vs.\ 3.33 tokens/s). All three H20 runs matched 400/400 sequences, matched TD CPU on 396/400, and scored 65.75\% EM (TITAN RTX matched TD CPU on 392/400 at 4B and 398/400 at 8B due to close logit ties). Here 15.115~s measures 400-turn \texttt{llama.cpp} generation, distinct from the 11.066~s OpenVINO 24-prompt remote mean (Table~\ref{tab:scale}).

\begin{table}[t]
\caption{MultiWOZ Quality, Memory, Startup, and Steady-State Wrapper Overhead on the 32~GB Intel TDX Instance (OpenVINO INT4)}\label{tab:scale}
\scriptsize
\renewcommand{\arraystretch}{1.04}
\begin{tabularx}{\columnwidth}{Yrrr}
\toprule
\textbf{Measure} & \textbf{Qwen3-0.6B} & \textbf{Qwen3-4B} & \textbf{Llama-3.1-8B} \\
\midrule
Token-identical MultiWOZ pairs & 400/400 & 400/400 & 400/400 \\
State EM / Slot micro-F1 (\%) & 60.25 / 91.65 & 64.25 / 92.92 & 66.50 / 93.21 \\
Key receipt to worker ready, mean (s) & 9.40 & 19.19 & 28.98 \\
Reboot to first ready worker, range (s) & 92--129 & 147--200 & 158--182 \\
Peak worker RSS / Guest use (GiB) & 5.77 / 6.90 & 18.77 / 21.61 & 24.36 / 29.32 \\
Mean remote request time (s) & 1.824 & 6.837 & 11.066 \\
Mean same-TD native worker time (s) & 1.593 & 6.567 & 10.803 \\
Mean remote $-$ native wrapper delta (s) & 0.231 & 0.270 & 0.263 \\
Median paired remote/native time ratio & 1.1410 & 1.0381 & 1.0257 \\
\bottomrule
\end{tabularx}
\end{table}

\begin{table}[t]
\caption{Summarization and Longer Minutes Generation on 32~GB TDX}\label{tab:summaries}
\scriptsize
\renewcommand{\arraystretch}{1.04}
\begin{tabularx}{\columnwidth}{Yrrr}
\toprule
\textbf{Benchmark \& Measure} & \textbf{Qwen3-0.6B} & \textbf{Qwen3-4B} & \textbf{Llama-3.1-8B} \\
\midrule
\multicolumn{4}{l}{\emph{DialogSum Short Summaries (100 test dialogues, 64-token cap)}} \\
Token-identical pairs / 64-cap hits & 100/100 / 8 & 100/100 / 2 & 100/100 / 1 \\
Mean ROUGE-L F1 / Mean tokens & 0.3941 / 36.95 & 0.4537 / 31.30 & 0.4534 / 33.56 \\
\midrule
\multicolumn{4}{l}{\emph{MeetingBank Council Minutes (32 agenda items, 512-token ceiling)}} \\
Token-identical pairs / 512-cap hits & 32/32 / 7 & 32/32 / 2 & 32/32 / 2 \\
Median content tokens (min--max) & 230.5 (108--512) & 142 (33--512) & 127 (34--512) \\
Outputs $>64$ / $>256$ content tokens & 32 / 14 & 30 / 5 & 31 / 2 \\
Max naturally terminated tokens & 483 & 336 & 172 \\
Mean ROUGE-L F1 & 0.2532 & 0.3849 & 0.4389 \\
\bottomrule
\end{tabularx}
\end{table}

On DialogSum (100/100) and MeetingBank (32/32), protected outputs matched native baselines token for token across all sizes (Table~\ref{tab:summaries}); 93/96 MeetingBank outputs exceeded 64 tokens and 82/93 stopped naturally below 512 (up to 483 at 0.6B), with 8B reaching 0.4389 ROUGE-L F1.

\subsection{Service-Wrapping Overhead, Memory, and Startup}
\begin{figure}[t]
\centering
\includegraphics[width=\columnwidth]{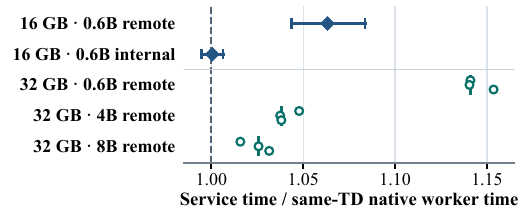}
\caption{Verified-byte service time relative to same-TD native worker time (16~GB: geometric means and 95\% bootstrap intervals over 12 boots; 32~GB: individual boot ratios and medians over 3 boots of 24 matched requests).}\label{fig:cost}
\end{figure}

Across 12 boots on 16~GB (0.6B), remote requests averaged 2.782~s vs.\ 2.615~s native (+0.167~s), yielding geometric-mean ratio 1.0632 ($95\%$ CI $[1.0436,1.0837]$; Fig.~\ref{fig:cost}) and internal-to-native ratio 1.0005 ($[0.9947,1.0063]$). On 32~GB (Table~\ref{tab:scale}), protocol wrapping added a near-constant 0.231--0.270~s (0.237--0.271~s outside the worker), lowering median ratios from 1.1410 (0.6B) to 1.0381 (4B) and 1.0257 (8B) within the same TD. At batch one, 8B fit in 32~GB at 24.36~GiB RSS and 29.32~GiB guest memory; key-to-ready time scaled from 9.40~s to 28.98~s (fastest on cache-warmed boot~2). Across six Secure Boot pairs, reboot-to-ready averaged 90.28~s vs.\ 87.26~s (+3.03~s, $[1.47,4.58]$), already present at first TLS handshake (+3.10~s, $[1.56,4.64]$) with unchanged post-key loading ($-0.01$~s, $[-0.09,0.07]$).

\section{Discussion and Limitations}
\textbf{Lifecycle and Residual Side Channels.}
Binding one base and private adapter per TD creates a clean authorization boundary across client requests; multi-adapter batching~\cite{slora,punica,dlora} requires per-tenant key and KV-cache isolation, and measured boot requires coordinated policy updates on guest changes. Attestation proves TCB and binary identity rather than bounding microarchitectural or traffic leakage: host instruction counting (\texttt{TDXdown}~\cite{tdxdown}), single-stepping/cache attacks (\texttt{TDXploit}~\cite{tdxploit}), and shared-resource timing~\cite{inteltdguidance} interact with prompt prefill, packet counts, and token access patterns.

\textbf{Cooperative CPU Serving vs.\ Confidential GPUs.}
While wrapping adds $\sim$0.25~s on 32~GB, 8B CPU latency is 11.066~s ($\sim$5 requests/min) vs.\ 0.414~s on CC-On H20 ($36.60\times$ throughput). Keeping tokenization, $\Delta W$, and decoding in one CPU TD avoids bounce buffers and multi-vendor quotes, suiting asynchronous cooperative tasks (state extraction, compliance review, minutes drafting) and interactive 0.6B serving, whereas interactive 8B serving needs confidential GPUs~\cite{loratee,slim,hoppertee} with joint host-TD and NRAS appraisal. Our protocol ports to AMD SEV-SNP~\cite{amdsnp} and Arm CCA~\cite{armrmm}; 29.32~GiB at 8B leaves tight 32~GB headroom, and ROUGE-L reflects lexical overlap rather than factuality.

\section{Conclusion}
We presented a bilateral Intel TDX LoRA inference service where an independent client and adapter owner verify the same measured service and TLS handshake key before releasing prompts or decryption keys, while an in-memory loader passes verified base, tokenizer, and decrypted adapter tensors directly to execution. Under Secure Boot and \texttt{UpToDate} TDX/QE policies, the service matched native outputs token for token across 0.6B, 4B, and 8B models on MultiWOZ, DialogSum, and MeetingBank, adding 0.231--0.270~s wrapper overhead on 32~GB TDs (1.0632 ratio on 16~GB 0.6B) and quantifying CPU-TD vs.\ CC-On H20 throughput boundaries (15.115~s vs.\ 0.414~s at 8B).

\ifanonymous\else
\section*{Acknowledgment}
Frozen LoRA adapters and two FP32 GGUF adapter exports are available at \url{https://github.com/mikelikeapex/tdx-lora-frozen-adapters/releases/tag/v1.0.0}. Per IEEE AI-content policy, the authors used OpenAI Codex (GPT-6, Nov.\ 2026) for benchmark tooling and experiment execution (Sections~IV--V), drafting (Sections~I--VII), and data-driven chart generation (Figs.~3--4), and Gemini in Antigravity (Nov.\ 2026) for English editing, formatting, and architecture/protocol diagram design (Figs.~1--2). The authors verified all outputs and take full responsibility for this work.
\fi

\bibliographystyle{IEEEtran}
\bibliography{main}

\end{document}